\documentclass[aip,showpacs,floatfix,amsmath,amssymb,amsfonts,unsortedaddress,superscriptaddress, reprint]{revtex4-1}

\newcommand{\figwidth}{ \linewidth}

\usepackage{graphicx}
\usepackage[normalem]{ulem}
\usepackage{color}
\usepackage{amsmath,amssymb}
\usepackage[T1]{fontenc}
\usepackage{siunitx}

\newcommand{\nhd}{NH$_3$}

\newcommand{\ndds}{ND$_3$ }
\newcommand{\nhds}{NH$_3$ }

\newcommand{\tpts}{$^3$P$_2$}

\newcommand{\tsos}{$^3$S$_1$}
\newcommand{\tso}{$^3$S$_1$ }

\newcommand{\beq}{\begin{equation}}
\newcommand{\eeq}{\end{equation}}

\newcommand{\chfs}{CHF$_3$ }
\newcommand{\chf}{CHF$_3$}

\begin{document}
\title{Importance of rotationally inelastic processes in low-energy Penning ionization of CHF$_3$}

\author{Justin Jankunas}
\affiliation{Institute for Chemical Sciences and Engineering, Ecole Polytechnique F\'ed\'erale de Lausanne (EPFL), 1015 Lausanne, Switzerland}
\author{Krzysztof Jachymski}
\affiliation{Faculty of Physics, University of Warsaw, Pasteura 5, 02-093 Warsaw, Poland}
\affiliation{Institute for Theoretical Physics III  and Center for Integrated Quantum Science and Technology, University of Stuttgart, Pfaffenwaldring 57, 70550 Stuttgart, Germany}
\author{Micha\l\ Hapka}
\affiliation{Faculty of Chemistry, University of Warsaw, Pasteura 1, 02-093 Warsaw, Poland}
\author{Andreas Osterwalder}
\affiliation{Institute for Chemical Sciences and Engineering, Ecole Polytechnique F\'ed\'erale de Lausanne (EPFL), 1015 Lausanne, Switzerland}
\email{andreas.osterwalder@epfl.ch}
\date{\today}
\begin{abstract}
Low energy reaction dynamics can strongly depend on the internal structure of the reactants.
The role of rotationally inelastic processes in cold collisions involving polyatomic molecules has not been explored so far.
Here we address this problem performing a merged-beam study of the He(\tsos)+\chfs Penning ionization reaction in a range of collision energies $E/k_B$=0.5--120 K. 
The experimental cross sections are compared with total reaction cross sections calculated within the framework of the quantum defect theory. 
We find that the broad range of collision energies combined with the relatively small rotational constants of \chfs makes rotationally inelastic collisions a crucial player in the total reaction dynamics. 
Quantitative agreement between theory and experiment is only obtained if the energy-dependent probability for rotational excitation is included in the calculations, in stark contrast to previous experiments where classical scaling laws were able to describe the results. 
\end{abstract}
\maketitle

Scattering processes between particles in the gas phase include elastic, inelastic, and reactive collisions.
In general, each of these processes has its own reaction rate, and a complete understanding of the collision dynamics requires knowledge of all these rates.
Because they all have an individual dependence on the collision energy, the complete characterization of molecular scattering events is a highly complex task, both theoretically and experimentally.
While in chemistry one is interested foremost in the reactive rates, studies of elastic and inelastic scattering processes are essential for understanding of molecular scattering and fundamental processes that determine chemical dynamics.~\cite{Nichols:2015hl,Eyles:2011hd,Brouard:2014et,vonZastrow:2014kg,Chefdeville:2013jc,Gubbels:2012js,Zuchowski:2009ws,Hutson:2007en}

In recent years, detailed investigations of scattering processes have attracted considerable interest in particular in the context of cold molecule formation and trapping.~\cite{Quemener:2012ua,Carr:2009ch}
Recent experimental developments have opened the route to investigations of molecular scattering in supersonic expansions at collision energies substantially below 10~K.~\cite{Henson:2012kr,2012PhRvL.109b3201C,Chefdeville:2013jc,LavertOfir:2014jea,shagam2015,Osterwalder:2015iq,Jankunas:2015fm,Jankunas:2014hg,Jankunas:2014jm,Bertsche:2014kl}
The collision energy depends on the relative velocity of the reactants.
In a molecular beams experiment this is determined not just by individual beam speeds but also by the angle between them.
Traditional crossed-beam measurements are performed at an angle of ninety degrees and, given the high speeds of supersonic expansions,\cite{Scoles:1988tm} cannot reach below 50 K.
This domain is currently accessible only by reducing the angle between the beams.~\cite{2012PhRvL.109b3201C,Chefdeville:2013jc}
To reach energies below 1 K, a complete superposition of the two beams is necessary which is the basis of the merged beams technique. 
Several studies in the past years have demonstrated the power of this approach, addressing different aspects of Penning ionization reactions in collisions of excited rare gas atoms with other atoms, diatomics or light polyatomic molecules.~\cite{Osterwalder:2015iq,Jankunas:2015fm,Jankunas:2014hg,Jankunas:2014jm,Bertsche:2014kl,LavertOfir:2014jea,Henson:2012kr,shagam2015}

Only a handful of low-energy collision studies of polyatomic molecules have been performed to date.~\cite{Jankunas:2014hg,Jankunas:2014jm,Bertsche:2014kl,Jankunas:2015fm} In general, higher mass and increasingly complex internal structure of the colliding molecules open more inelastic and reactive channels.
Therefore, the complete description of a cold chemical reaction must account for all possible reaction routes and also include inelastic collisions.

\begin{figure}
\includegraphics[width=\figwidth]{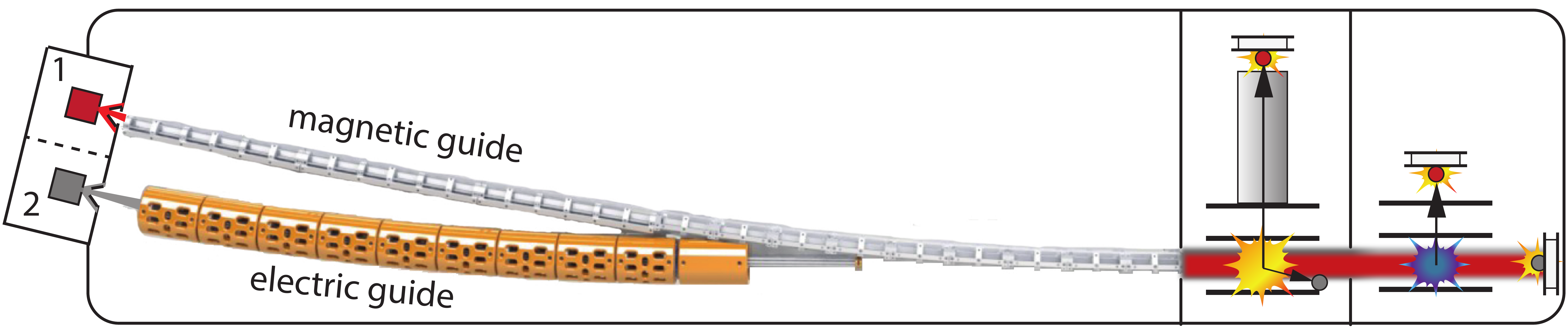}
\caption{\label{expt}The experimental setup used in the present merged beams study. Two supersonic expansions (labeled 1 and 2) are injected into a bent magnetic and electric guide, respectively. The geometry of these guides produces two beams moving along the same axis. Appropriate timing of the valves and choice of expansion conditions produces two overlapping packets of reactants at well defined collision energy. Reaction products are detected in a pulsed time-of-flight mass spectrometer, and beam densities are constantly monitored on separate detectors. The entire setup is housed in a four-fold differentially pumped high-vacuum chamber.\cite{Osterwalder:2015iq}}
\end{figure}
Here, we investigate the contribution of rotationally inelastic collisions to the total reaction cross section as a function of collision energy.
Using the experimental setup shown in Fig. \ref{expt} and described in the methods section below, we produce and superpose pulsed supersonic expansions of He(\tsos) and \chf.\cite{Osterwalder:2015iq}
The collision energy is given as $E_{coll}=\frac{\mu}{2}(v_1-v_2)^2$, where $\mu$ is the reduced mass and the $v_i$ are the beam velocities which are controlled experimentally via the choice of valve temperature and carrier gas.
In the current reaction we tune the collision energy between 500 mK and 120 K, thus covering almost three orders of magnitude in energy.
Because the ionization energy of \chfs is less than the internal energy of He(\tsos) such collisions can lead to Penning ionization, accompanied by dissociation.
Specifically, the following principal reactive routes are energetically accessible:\cite{Torres:2002jp}
\begin{eqnarray*}
\mathrm{He}(^3\mathrm{S}_1)+\mathrm{CHF}_3	&\longrightarrow&\mathrm{He}(^1\mathrm{S})+		\mathrm{CF}_3^++\mathrm{H}+e^-, \\
										&\longrightarrow&\mathrm{He}(^1\mathrm{S})+		\mathrm{CHF}_2^++\mathrm{F}+e^-, \\
										&\longrightarrow&\mathrm{He}(^1\mathrm{S})+		\mathrm{CF}^++\mathrm{HF}+\mathrm{F}+e^-.
\end{eqnarray*}
CHF$_3^+$ itself is not observed, presumably because all of the accessible states readily (pre-)dissociate.\cite{Torres:2002jp,Parkes:2006iv}
The ionized molecules are detected in a time-of-flight mass spectrometer, and product ions CF$^+$, CHF$_2^+$, and CF$_3^+$ are counted individually to monitor reaction channels separately as a function of collision energy. 

In the present work we probe a wide energy range, from low collision energies where no channels for rotationally inelastic collisions are open, to energies where rotationally inelastic collisions become possible and, in fact, dominate the collision dynamics. 
By recording the probability of reactive collisions and comparing it with the calculated total collision cross section we obtain -- in an indirect fashion -- the probability for inelastic collisions.

\section{Theoretical model}
Theoretical modeling of collisions between polyatomic molecules and electronically excited atoms is an extremely challenging task. 
Rigorous treatment of the Penning ionization process involving a molecule would require obtaining the relevant potential energy surfaces with a precision that is currently beyond reach, even using state-of-the-art numerical procedures. However, effective models can be constructed to capture the essential features of the problem.~\cite{Siska:1993cf,Hapka:13,Pawlak2015} 
In the current study, the rotational structure of \chfs can be expected to have strong importance, as its rotational constants are comparable to the experimental energy scales.
This leads to the opening of new accessible channels with increasing energy, thus increasing the complexity of the problem even further.
Here we do not aim for a full scattering calculation involving all the relevant channels, but only to construct an effective model to explain the observed rate of ionization events.
To this end, we make use of the description based on quantum defect theory (QDT), which has been successfully applied before to describe Penning ionization of the NH$_3$ molecule.~\cite{Jankunas:2014hg,Jankunas:2014jm} 
The model relies on the knowledge of the interaction potential at long range and an effective parametrization of the short-range dynamics (see refs. ~\onlinecite{Jachymski2013,Jankunas:2014hg} for details).
The necessary parameters of the model are the short-range reaction probability $P_{\rm re}$ and the short-range phase $\phi$ that sets the scattering length of the entrance channel potential and controls the positions of shape resonances. 
Both quantities can in principle depend on the collision energy and partial wave quantum number, and are thus sensitive to the presence of rotationally excited states.

A first possible effect are densely spaced resonances that may occur due to coupling to bound states supported by the closed channels. At a resonance, the short range phase varies rapidly.
A second consequence is that multiple open exit channels corresponding to rotational excitations appear as the collision energy grows, so the probability for reaction should decrease.
It is possible to estimate the density of states affecting the reaction just by knowing the rotational constants and using simplified potential energy curves.\cite{Mayle2012,Mayle2013}
but only account for their impact on our QDT parameters $P_{\rm re}$ and $\phi$. 
For simplicity, we assume that all the possible inelastic processes that lead to rotational excitation have similar probability.
If this is the case, then the probability for a reactive collision can be described as
\begin{equation}
P_{\rm re}(E)=\frac{1}{1/P_0+\gamma \mathcal{N}(E)},
\end{equation}
where $P_0$ is the reaction probability at zero energy, which was fitted to $P_0$=0.1.
$\mathcal{N}$ counts the number of energetically accessible rotational states and $\gamma$ is a~fitting parameter describing the branching ratio between reactive and inelastic events, assumed to be constant over the entire range of energies studied here. 
For each calculation we randomly choose energies at which $\pi$-shifts of the short-range phase occur, resulting in resonances with density approximately governed by the expected density of states. 
Naturally, the distribution of these resonances depends on the level structure of the molecule, and thus on both potential and rotational constants.

In QDT one can then perform scattering calculations using $P_{\rm re}$ and $\phi$ to construct short-range boundary conditions. The last thing that is needed is the interaction potential at long range, which is dominated by van der Waals attraction.
We have obtained the leading coefficients for dispersion $C_{6,{\rm disp}}^{00}=350.9$ a.u. and induction $C_{6,{\rm ind}}^{00}=105.4$ a.u. using symmetry-adapted perturbation theory (SAPT(UHF)) \cite{Hapka:12} and suitable asymptotic expressions of the multipole expansion, respectively~\cite{Schmuttenmaer:91}. 
Moreover, in order to estimate the well depth of the molecular complex we also performed SAPT(UHF) calculations up to the second order in intermolecular interaction operator. The global minimum of 72.64 cm$^{-1}$ corresponding to the collinear He(\tsos)-HCF$_3$ geometry was found.

\section{Results}

\begin{figure}
\includegraphics[width=\figwidth]{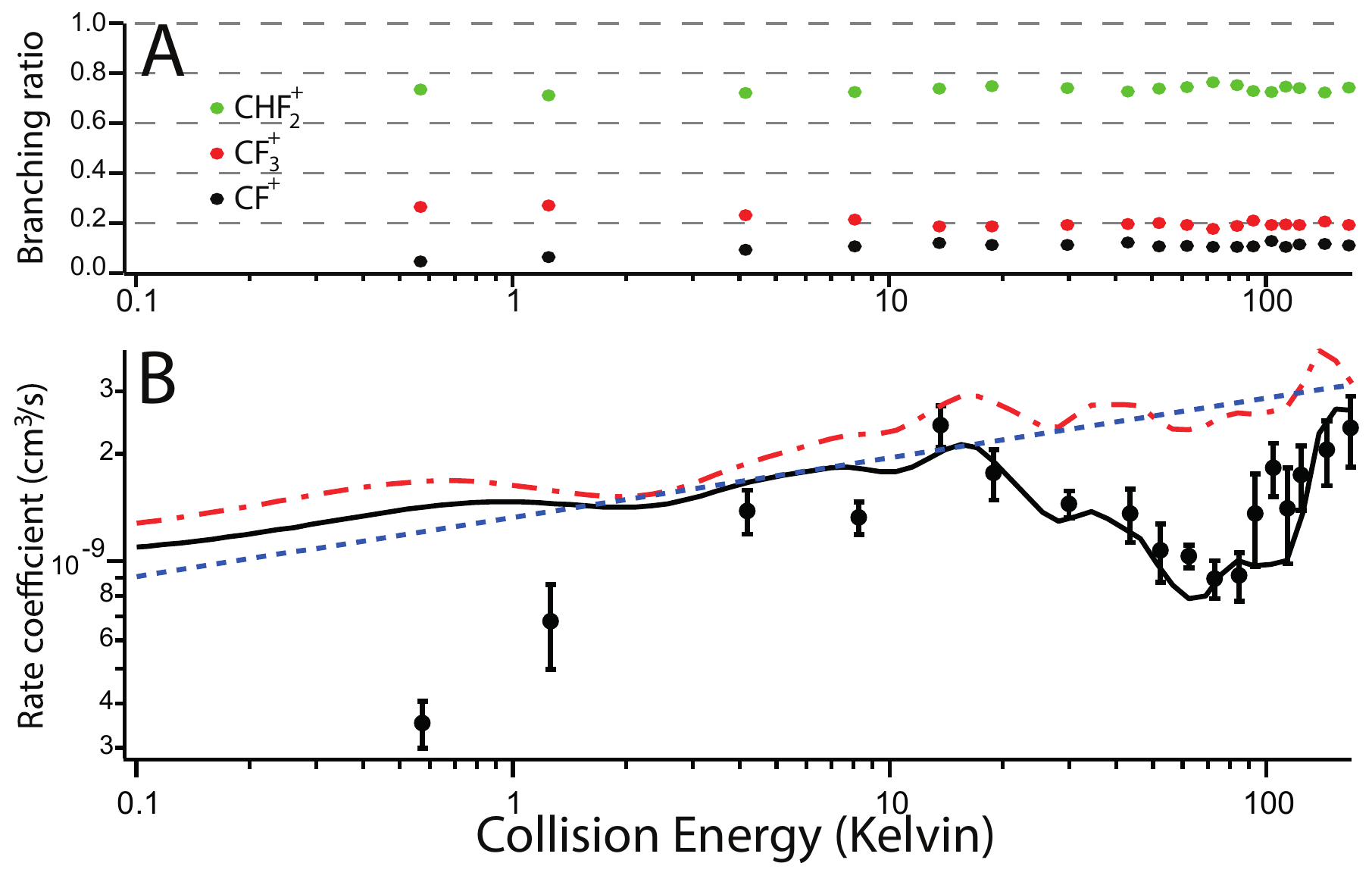}
\caption{\label{data}Experimental and principal theoretical results. A) Experimental branching ratios for the three fragments CF$^+$, CHF$_2^+$, and CF$_3^+$ formed in the dissociative Penning ionization of \chfs by metastable helium. B) Experimentally determined reaction rate coefficients for the He(\tso)+\chfs reaction; black solid line: reaction rate constants resulting from QDT calculations; red dash-dotted line: results of a calculation that does not include inelastic collisions; blue dashed line: prediction of classical Langevin capture model.}
\end{figure}

Fig. \ref{data} shows the most important results of our study. As seen in Fig. \ref{data}A, the branching ratio between the three accessible reaction channels is nearly constant over the entire energy range covered here.
The experimental rate coefficient for formation of CF$^+$ is shown in Fig. \ref{data}B as black dots (the experiment does not provide absolute rate coefficients; the scale is determined through the calculations).
The red dash-dotted curve in that panel is the rate coefficient from our QDT model calculations in the absence of inelastic collisions.
The dashed blue line is the prediction from a pure Langevin capture model that for van der Waals interactions predicts the reaction rate constant to scale with $E_{\rm coll}^{1/6}$. Clearly, our results do not follow any power law behavior.
The impact of rotational excitations is visible above 10 K where the red curve disctinctly overestimates the rates.
Satisfactory agreement is achieved only when inelastic collisions are included, as shown by the black solid line in Fig \ref{data}B. 
The minimum of the reactive rate coefficient is located at the energy corresponding to the binding energy of the excited complex, and its presence is independent of inelastic collisions.~\cite{Siska:1993cf,Haberland:1981ew} The position of this minimum obtained from QDT calculations closely matches the experimental result.

\begin{figure}
\includegraphics[width=\figwidth]{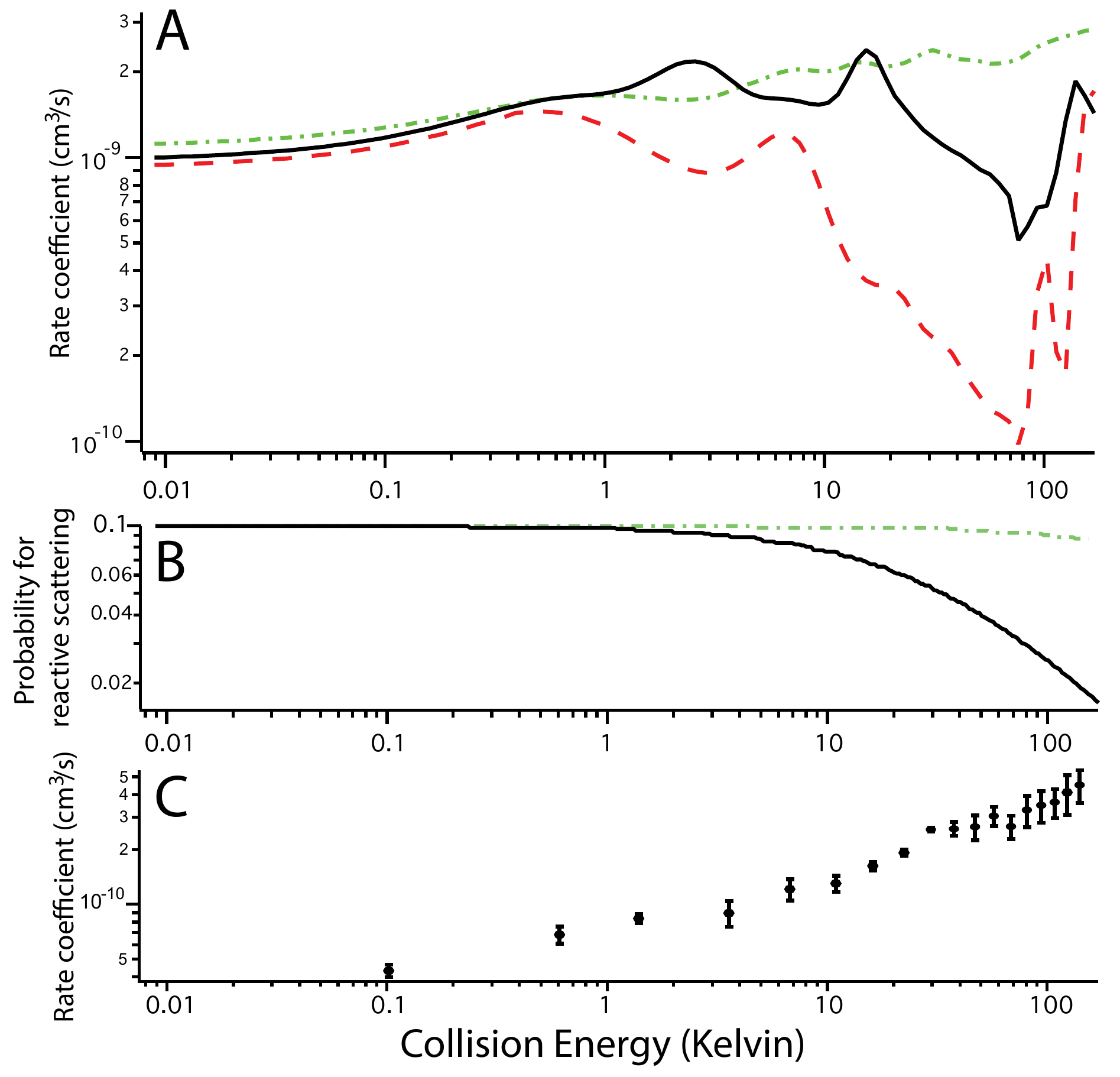}
\caption{\label{model}Supporting calculations. A) comparison of different calculations assuming different rotational constants. Solid black line: calculation for \chf. Dash-dotted green line: rotational constants of \nhd. Dashed red line: rotational constants ten times smaller than \chf. B) probability for reactive collisions, assuming inelastic reaction probability proportional to the total number of accessible states. Green and black lines show the results for \nhds and \chf, respectively. C) experimental rate coefficients determined for the Ne(\tpts)+\ndds reaction.\cite{Jankunas:2014hg} In contrast to the present reaction there is no signature of suppression of reactive rates by inelastic collisions.}
\end{figure}

\section{Discussion}
In order to further explore the features of our model we show, in Fig.~\ref{model}A, results of calculations in which we changed the rotational constants of the molecule while keeping the interaction potential of the He(\tsos)-\chfs complex. 
The dash-dotted green and solid black lines were computed assuming rotational constants for \nhds and \chf, respectively, and the dashed red line corresponds to a molecule with rotational constants ten times smaller than \chf. 
Because of the large rotational constant of \nhds most inelastic collision channels remain closed and the green curve in Fig. \ref{model}A only slightly differs from the red curve in Fig.~\ref{data}B at high energies. 
Moreover, we assume that the probability for rotationally inelastic scattering is directly proportional to the number of accessible channels. 
In consequence, \chfs with rotational constants\cite{Anonymous:TYz8Fa5J} of $A = B = 0.3452$ cm$^{-1}$ and $C =0.18925$ cm$^{-1}$ should experience more rapid drop in the probability for reactive scattering than \nhds with rotational constants\cite{Anonymous:TYz8Fa5J} of $A = B = 9.4443$ cm$^{-1}$ and $C =6.1960$ cm$^{-1}$. 
This effect is shown in Fig.~\ref{model}B. 
While the direct proportionality assumed in our model is only an approximation,\cite{Gubbels:2012js} we stress that it it is this inclusion of inelastic scattering that provided satisfactory agreement with the experimental data.
Finally, for comparison, Fig. \ref{model}C shows the experimental results from our study of the Ne*+\ndds reaction.\cite{Jankunas:2014hg}
Both experiment and theory confirm that for light molecules with large rotational constants it is not necessary to explicitly include inelastic scattering channels.

Our simplified theoretical treatment allows only to estimate the Penning ionization rates and is not suitable for calculating the rates for production of rotationally excited states. However, a full multichannel model would inevitably contain a large number of free parameters needed to describe couplings to open channels. We stress that due to experimental energy resolution and very large number of channels, the detailed behaviour of each single inelastic process would anyway be washed out and change the ionization rates only slightly. As a result, our QDT model works well despite its crudeness. A detailed theoretical study of competing reactive and inelastic processes in \chfs would allow to understand the system more deeply, but for now remains an open challenge.

\section{Conclusions}
We have demonstrated that rotationally inelastic collisions may drastically reduce Penning ionization rates, even in the low-energies regime of a few Kelvin. This effect is related to the magnitude of the rotational constant which appears to be critical. When the rotational constant is low, a considerable number of inelastic collision channels may be open and suppress reactivity. Indeed, suppressed reactivity at high energies has been observed in the case of CHF$_3$, but not in any of the lighter systems.~\cite{Jankunas:2015fm,Jankunas:2014hg,Jankunas:2014jm,Bertsche:2014kl,shagam2015,LavertOfir:2014jea,Henson:2012kr}
The principal difference between \chfs and CH$_3$F or ammonia are larger mass and smaller rotational constants. This leads to a different appearance of the energy dependent reaction rate coefficient. Based on these results we suggest that this is a general property of molecular collisions: collisions of heavier molecules will be largely determined by inelastic scattering, even at the lowest temperatures. This striking result means that classical scaling laws that are known to work for barrierless reactions of light molecules cannot be applied to polyatomic molecules. 

\section{Acknowledgments}
This work was supported by Foundation for Polish Science within the START program, Alexander von Humboldt Foundation, Swiss Science Foundation (Grant No. PP0022-119081), and by EPFL.


%

\end{document}